\documentclass[prl,reprint,twocolumn,superscriptaddress,noshowpacs,notitlepage,longbibliography]{revtex4-1}%

\usepackage{graphicx,bm,times}
\usepackage{amsmath}
\usepackage{amsfonts}
\usepackage{amssymb}
\usepackage{color}
\usepackage{hyperref}
\hypersetup{
	colorlinks = true,
	allcolors = {blue}
}

\begin{document}

\title{Shifts and Splittings of the Hole Bands in the Nematic Phase of FeSe}

\author{Matthew D. Watson}
\affiliation{Diamond Light Source, Harwell Campus, Didcot OX11 0DE, United Kingdom}

\author{Amir A. Haghighirad}
\affiliation{Clarendon Laboratory, Department of Physics,
	University of Oxford, Parks Road, Oxford OX1 3PU, United Kingdom}

\author{Hitoshi Takita}
\affiliation{Graduate School of Science, Hiroshima University, 1-3-1 Kagamiyama, Higashi-Hiroshima 739-8526, Japan}

\author{Wumiti Mansur}
\affiliation{Graduate School of Science, Hiroshima University, 1-3-1 Kagamiyama, Higashi-Hiroshima 739-8526, Japan}

\author{Hideaki Iwasawa}
\affiliation{Hiroshima Synchrotron Radiation Center, Hiroshima University, 2-313 Kagamiyama, Higashi-Hiroshima 739-0046, Japan}
\affiliation{Diamond Light Source, Harwell Campus, Didcot OX11 0DE, United Kingdom}

\author{Eike F. Schwier}
\affiliation{Hiroshima Synchrotron Radiation Center, Hiroshima University, 2-313 Kagamiyama, Higashi-Hiroshima 739-0046, Japan}

\author{Akihiro Ino}
\affiliation{Graduate School of Science, Hiroshima University, , 1-3-1 Kagamiyama, Higashi-Hiroshima 739-8526, Japan}
\affiliation{Hiroshima Synchrotron Radiation Center, Hiroshima University, 2-313 Kagamiyama, Higashi-Hiroshima 739-0046, Japan}

\author{Moritz Hoesch}
\affiliation{Diamond Light Source, Harwell Campus, Didcot OX11 0DE, United Kingdom}
\affiliation{Hiroshima Synchrotron Radiation Center, Hiroshima University, 2-313 Kagamiyama, Higashi-Hiroshima 739-0046, Japan}

\begin{abstract}
We report a high-resolution laser-based angle-resolved photoemission spectroscopy (laser-ARPES) study of single crystals of FeSe, focusing on the temperature-dependence of the hole-like bands around the ${\rm \Gamma}$ point. As the system cools through the tetragonal-orthorhombic ``nematic" structural transition at 90~K, the splitting of the $d_{xz}$/$d_{yz}$ bands is observed to increase by a magnitude of 13 meV. Moreover, the onset of a $\sim$10 meV downward shift of the $d_{xy}$ band is also at 90~K. These measurements provide clarity on the nature, magnitude and temperature-dependence of the band shifts at the ${\rm \Gamma}$ point in the nematic phase of FeSe. 
\end{abstract}

\date{\today}
\maketitle


{\it Introduction.-}
The drive to understand the origin of unconventional superconductivity in the iron-based superconductors has led to a focus on the ordered phases also found in their phase diagrams. Amongst these, the so-called ``nematic phase" has been intensively studied, but remains poorly understood. The nematic phase is characterised by a tetragonal to orthorhombic phase transition of the lattice, and there is a general consensus that it is driven by electronic degrees of freedom; however the relative importance of spin and orbital degrees of freedom is still debated \cite{Fernandes2014,Glasbrenner2015,Chubukov2015}. Recently, FeSe has become a focal point of this discussion due to its unique properties: it exhibits a tetragonal-orthorhombic transition at 90~K but no magnetic ordering at any temperature, in addition to fascinating phase diagrams under pressure \cite{Sun2016} and chemical substitution \cite{Watson2015c}.  It was suggested that FeSe might be a system where orbital ordering, and not magnetic interactions, may play the dominant role in the nematic order \cite{Bohmer2014}, and various orbital ordering scenarios have been proposed \cite{Jiang2016,Watson2015a,Watson2016a,Zhang2015,Xing2016_arxiv,Kreisel2016_arxiv}, while on the other hand the presence of strong paramagnons \cite{Rahn2015,Wang2016} and possible magnetic frustration \cite{Glasbrenner2015,Wang2016b} would suggest that magnetic interactions may yet play an important role.

Key to the resolution of this debate are details of the evolution of the band structure of FeSe, for which high-resolution angle-resolved photoemission spectroscopy (ARPES) measurements can provide a unique experimental insight. In principle, by observing splittings and band shifts at different points in the Brillouin zone and making comparison to measurements in the higher temperature tetragonal phase, the form of the orbital order parameter in the low-temperature phase could be completely constrained. However in practice there has been a significant controversy surrounding the interpretation of ARPES measurements of FeSe. This is particularly the case for the electron pockets at the $\bar{\rm M}$ point, where a large 50 meV energy scale has been associated with the lifting of $d_{xz}-d_{yz}$ degeneracy in several papers \cite{Fanfarillo2016,Nakayama2014,Shimojima2014,Zhang2015}, but interpreted as separate $d_{xz/yz}$ and $d_{xy}$ bands in other recent works \cite{Watson2016a,Fedorov2016}. For the hole pockets around the $\bar{\rm \Gamma}$ point, the interpretation of the data ought to be more straightforward, however there are also conflicting reports here \cite{Zhang2015,Watson2015a,Watson2016a}. The resulting confusion over which order parameters are consistent with the experimental data has hampered progress towards the robust identification of (the orbital component of) the nematic order parameter in FeSe. There is therefore a strong motivation to revisit the problem of the temperature-dependence of the hole pockets of FeSe in detail, which is a good match to the capabilities of laser-ARPES. 

In this paper we present a study of the evolution of the hole-like bands of single crystals of FeSe using high-resolution laser-ARPES. We perform a fitting analysis of the data to extract band positions at the $\bar{\rm \Gamma}$ point as a function of temperature. We observe that the $d_{xz}$/$d_{yz}$ hole bands are split due to spin-orbit coupling at high temperature, but the splitting has a marked increase at low temperature which onsets exactly with the structural transition at 90~K. Furthermore we report a downward band shift of the $d_{xy}$ hole band at the $\bar{\rm \Gamma}$ point which accompanies the nematic ordering. Our results clarify the nature, magnitude and temperature-dependence of the band shifts at the $\bar{\rm \Gamma}$ point in the nematic phase of FeSe, and we discuss how our results constrain, support, or rule out the various orbital order parameters suggested in the literature.

\begin{figure}[t!]
	\centering
	\includegraphics[width=\linewidth]{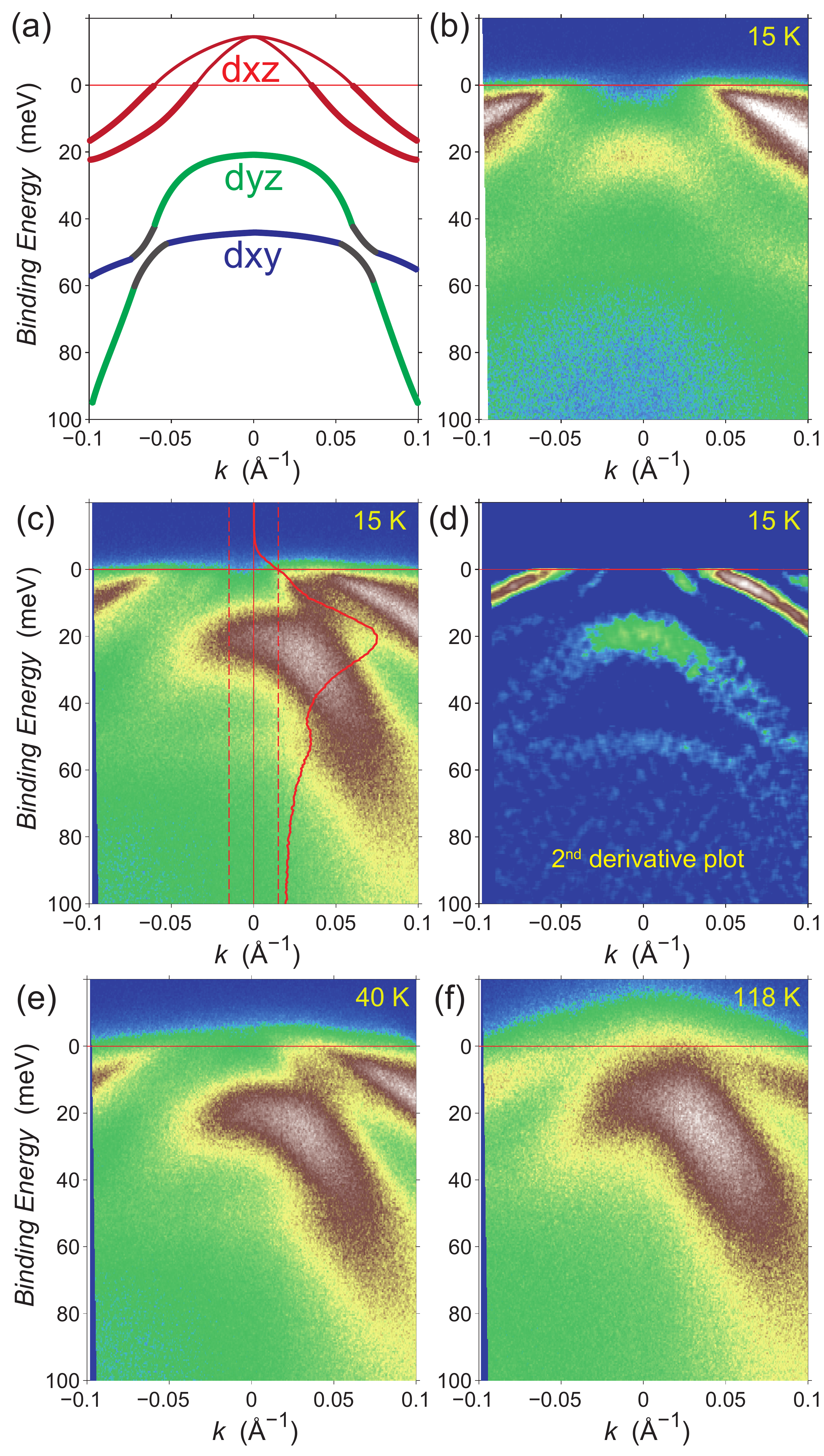}
	\caption[Fig1]{a) Schematic band structure of FeSe, for a slice through the $\bar{\rm \Gamma}$ point in the $\bar{\rm M}\bar{\rm \Gamma}\bar{\rm M}$ direction. b) Laser ARPES measurement in $s$ polarisation. c) Measurement obtained in a mixed polarisation. Dashed red lines show the integration window ($\pm$0.015 \AA$^{-1}$) for the EDCs used for the fitting analysis. d) Second-derivative plot of c). e,f) Higher temperature data obtained in the same geometry as c).}
	\label{fig:fig1}
\end{figure}

{\it Methods.-}
Single crystals of FeSe were grown by the vapor-transport method
\cite{Watson2015a}. Laser-ARPES measurements were performed at Synchrotron Radiation Center (HSRC), Hiroshima using a photon energy of 6.268 eV \footnote{H. Iwasawa, in preparation}. Fresh surfaces were prepared by cleavage at $T = 20$~K. The high-symmetry measurement geometry through normal emission (Fig.~\ref{fig:fig1}) was determined by a rotation scan of the sample tilt, after choosing a suitable part of the sample which gave a homogeneous signal over a $> 100\times 100~\mu$m$^2$ area. The temperature was raised from $T=15$~K in steps; the sample position was kept constant (within $\sim 20~\mu$m) with small positional adjustments based on a survey of two optical cameras to compensate for the thermal expansion of the sample manipulator. The sample quality was checked by obtaining identical data after cooling back down to $T=19$~K.

\begin{figure*}
	\centering
	\includegraphics[width=\linewidth]{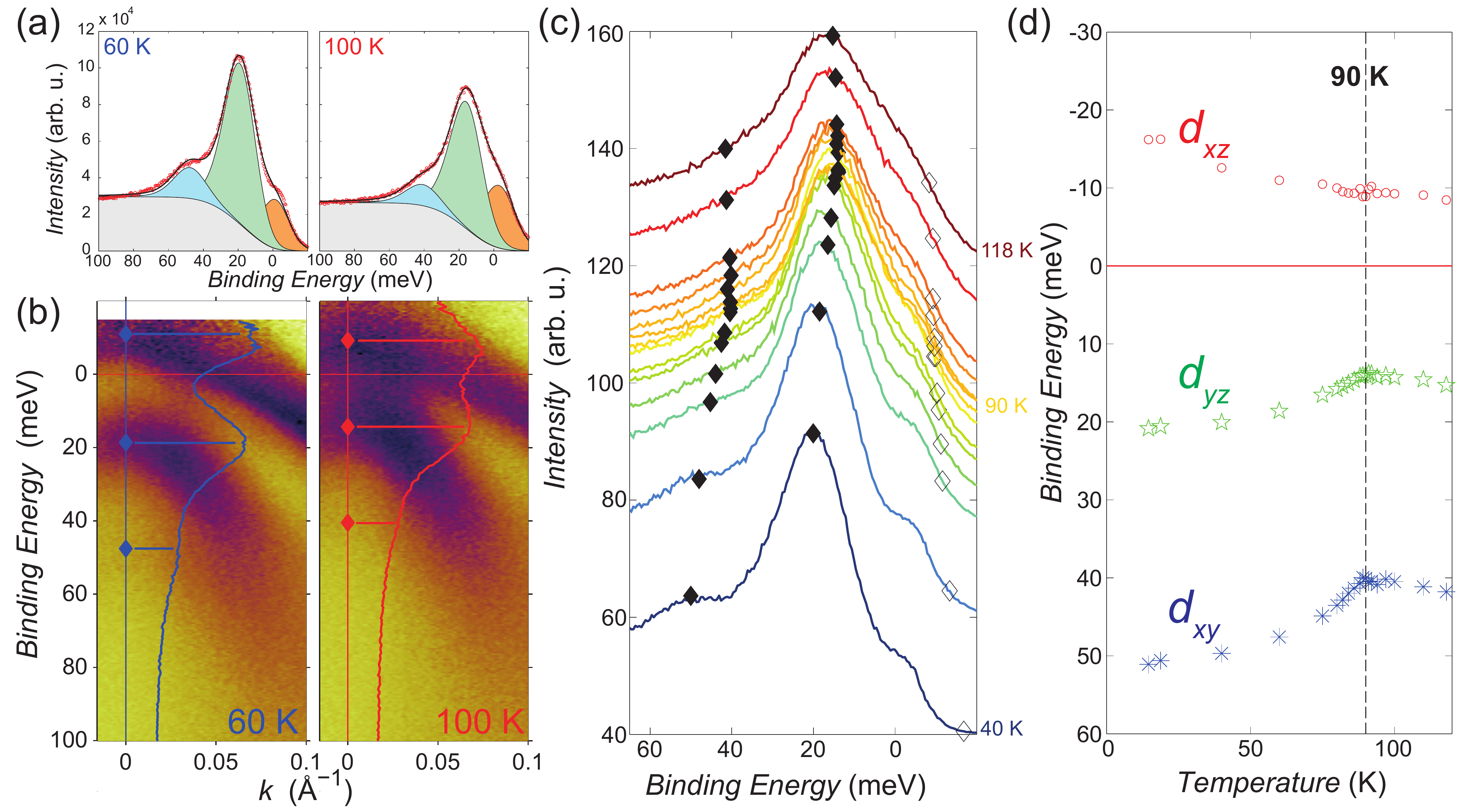}
	\caption[Fig2]{a) Selected EDCs, showing the fit and individual peak lineshapes and background components. b) Comparison of ARPES data at 60~K and 100~K; data are divided by the Fermi function to reveal the band dispersions above $E_F$. Extracted band positions are also marked. c) Selected EDCs stacked according to temperature. d) Temperature-dependence of band positions at the $\bar{\rm \Gamma}$ point from the fitting analysis.}
	\label{fig:fig2}
\end{figure*}

{\it Results.-} 
In Fig.~\ref{fig:fig1}a), we show the sketch of the hole bands of FeSe \cite{Watson2015a,Suzuki2015}, focusing on the energy and momentum region accessible by our laser-ARPES set-up. The bands are labelled by their dominant orbital characters in this high-symmetry cut, consistent with other reports, although in reality the orbital characters on the bands are mixed, especially near anticrossings. Note that this schematic diagram refers to the low temperature phase in a twinned sample, where a superposition of both orthorhombic domains is observed in experiment, so that both the inner and outer branches of the elliptical $d_{xz}$ pocket are observed. In principle the inner $d_{yz}$ band should also display split dispersions away from $\bar{\rm \Gamma}$, however this detail is yet to be clearly resolved. For the experimental ARPES measurement shown in Fig.~\ref{fig:fig1}b) using $s$ polarisation, the spectrum is dominated by the outer hole band with primarily $d_{xz}$ character. The outermost $k_F$ value is estimated to be 0.06(1) \AA$^{-1}$. By comparison with previous photon-energy dependent studies \cite{Watson2015a}, this implies that our laser-ARPES is measuring at an effective $k_z$ close to the 3D $\Gamma$ point.

For the purposes of this study, a strong intensity of the inner $d_{yz}$ band and a resolvable $d_{xy}$ peak at the $\bar{\rm \Gamma}$ point were desirable. In order to meet this condition, we used a mixed polarisation, shown in Fig.~\ref{fig:fig1}c). Although this spectrum is asymmetric, the Energy Dispersion Curve (EDC) at $\bar{\rm \Gamma}$ is more suitable for the extraction of band positions from fitting the EDCs. In Fig.~\ref{fig:fig1}d) we plot the second derivative of this mixed polarisation measurement, which shows traces of all the expected bands.

We now perform a fitting analysis of the temperature-dependent EDCs at the $\bar{\rm \Gamma}$ point, using a model with as few free parameters as possible. We chose to fit the data with three peaks representing the three band positions, a background function to account for secondary electrons and the Fermi function, all convoluted with the experimental resolution. For the peaks, we chose pseudo-Voigt functions \footnote{We used psuedo-Voigt functions, that is a linear sum of a Gaussian and a Lorentzian lineshape, in our case fixed to have equal amplitude for both components.}, although the results do not depend strongly on this choice. In order to minimize the number of free parameters we constrained all the peaks to have equal width. Moreover the peak amplitude of the $d_{xz}$ band was fixed to be equal to its relaxed value at high temperature. A simplistic approximation to the secondary electron background is incorporated with an error function which is centered on the central peak with the largest amplitude, and also has the same width. Thus the model has 7 free parameters, including the 3 peak positions. 

In Fig.~\ref{fig:fig2}a) we show representative fits to the experimental data, where it can be seen that the fit matches the data well while also giving physically reasonable peak and background lineshapes. It should be noted that the position of the $d_{xz}$ band above $E_F$ has a larger degree of uncertainty, especially at low temperatures, as the peak only partially contributes to the spectrum before the Fermi cut-off \footnote{By fixing the magnitude of this peak at high temperatures and linking its width to the $d_{yz}$ peak, we ensure that the temperature-dependence of this peak position can be reliably interpreted}. In Fig.~\ref{fig:fig2}b) we show ARPES data divided by the Fermi function to reveal the features just above the Fermi level. As can be seen, the band positions from the peak fitting all agree with features that can be observed directly in the data, giving confidence in our methodology. In Fig.~\ref{fig:fig2}c), we show a selection of integrated EDCs; it can be readily observed here that the peak positions in the occupied states do not evolve smoothly as a function of temperature, but have a kink at 90~K. The extracted peak positions as a function of temperature, which are the central result of this paper, are shown in Fig.~\ref{fig:fig2}d). 

In the high temperature tetragonal phase, the $d_{xz}$ and $d_{yz}$ bands are split by spin-orbit coupling only \cite{Watson2015a,Fernandes2014b}. Thus we estimate the magnitude of the spin-orbit induced splitting at the $\Gamma$ point to be 23.5~meV, similar to previous reports \cite{Watson2015a,Borisenko2015}. Below 90~K, an increased splitting of the bands is expected if the nematic order parameter has a symmetry breaking term such as $\frac{\Delta{}}{2}(n_{yz}-n_{xz})$ taking a non-zero value at $\mathbf{k}$=0 \cite{Fernandes2014b}. Thus our observation of an order parameter-like increase of the separation of $d_{xz}$ and $d_{yz}$ bands may be considered a direct signature of symmetry-breaking, since the spin-orbit term is expected to be temperature-independent. By low temperatures, the magnitude of the splitting reaches 37.5~meV, an increase of the band separation by 14~meV. We note that this actually implies a significantly larger value of $\Delta{}\sim{}\sqrt{(37.5^2-23.5^2)}\sim{}$29~meV, as the nematic splitting adds in quadrature, not linearly, with the gap due to spin-orbit coupling \cite{Fernandes2014b,Xing2016_arxiv}. 

 We also observe that the binding energy of the $d_{xy}$ band increases by $\sim$ 10 meV below 90~K. This effect has not been previously reported at the $\bar{\rm \Gamma}$ point, although it has been observed that the saddle point of the $d_{xy}$ bands at the $\bar{\rm{M}}$ point moves to higher binding energies by about 10 meV below 90~K \cite{Watson2016a}. Since the magnitude and direction of these band shifts in different regions of the Brillouin zone are in agreement, it seems that the $d_{xy}$ band shift is momentum-independent. Two interpretations of this feature are possible. The first interpretation is that the $d_{xy}$ band shift is an intrinsic feature of nematic order. However, most studies consider the nematic order parameter to originate in the $d_{xz/yz}$ orbitals which are globally degenerate in the tetragonal phase, and furthermore a rigid shift seems an unlikely consequence of symmetry-breaking order. The second interpretation is that the momentum-independent shift of the binding energy of the $d_{xy}$ band is not fundamentally related to the nematic order parameter, but rather is a manifestation of a shift of the chemical potential which is necessary for charge conservation when the nematic order parameter sets in and shifts the $d_{xz/yz}$ bands near the Fermi level. However differences between these two interpretations are subtle, and a combination of both effects is also possible. The uniform downward trend of all band positions at the $\bar{\rm \Gamma}$ point above 90~K is not an artefact, but is a real effect related to a gradual temperature dependence of the chemical potential in FeSe in the tetragonal phase \footnote{L. C. Rhodes, in preparation}.


Our results have important implications for the discussion of the nematic order parameter. The observation of a marked increase in the splitting of the $d_{xz}/d_{yz}$ bands at the $\bar{\rm \Gamma}$ point below 90~K rules out the scenario of $d$-bond nematic order \cite{Zhang2015,Jiang2016}. The observations in this work alone could be consistent with ferro-orbital ordering, however that scenario has already been conclusively eliminated by detwinned studies of FeSe \cite{Suzuki2015}. The rarely-considered extended $s$-wave bond nematic order \cite{Jiang2016} would be a possible explanation of our data set at the $\bar{\rm \Gamma}$ point, but this order parameter takes zero value at the $\bar{\rm M}$ point, in contrast with the significant band shifts detected experimentally \cite{Watson2016a}. Therefore in our understanding, the ``unidirectional nematic bond order" (which may be considered as a variant of the extended $s$-wave bond nematic order) recently proposed by some of us \cite{Watson2016a} is the only order parameter which is broadly consistent with all experimental data. There is not yet consensus on these issues \cite{Fanfarillo2016,Fedorov2016,Kreisel2016_arxiv}, however we hope that this study will provide much-needed clarity on the behavior of the hole bands of FeSe in the nematic phase. 

{\it Conclusion.-}
We have performed a detailed laser-ARPES study of the hole pockets of FeSe, revealing the onset of an increased splitting of $d_{xz}/d_{yz}$ bands below 90~K which is associated with a symmetry-breaking orbital ordering. In addition the $d_{xy}$ band also shifts to higher binding energies below 90~K. These results help to constrain models of ordering in the nematic phase of FeSe.

\begin{acknowledgments}
\section{Acknowledgments}
We thank A.~I.~Coldea, K. Shimada, T.~K.~Kim and L.~C.~Rhodes for insightful discussions. We wish to thank H. Namatame and M. Taniguchi for discussions and leadership that led to the creation of the laser-ARPES instrument. The high quality of sample handling instrumentation developed by Y. Aiura contributed to the result of this study. Beam time at the HSRC was granted under proposal number 16BU008. A.A.H. acknowledges the financial support of the Oxford Quantum Materials Platform Grant (EP/M020517/1).

\end{acknowledgments}
\bibliography{FeSe_laser_bib}
%

\end{document}